\documentclass[apj]{emulateapj}
\submitted{ApJ, accepted 28/05/2016}

\usepackage{amsmath}
\usepackage{amsfonts}
\usepackage{amsthm}
\usepackage{amssymb}
\usepackage{graphicx}
\usepackage{mathrsfs}
\usepackage{hyperref}
\usepackage{xspace}

\newcommand{\be}{\begin{equation}}
\newcommand{\ee}{\end{equation}}
\newcommand{\eq}[1]{Equation~(\ref{eq_#1})}
\newcommand{\fig}[1]{Figure~\ref{fig_#1}}
\renewcommand{\S}{Section~}

\newcommand{\kms}{{\rm km\,s^{-1}}}
\newcommand{\msun}{{\rm M}_{\odot}}

\newcommand{\ha}{H{\sc\,i}\xspace}
\newcommand{\hm}{H$_2$\xspace}
\newcommand{\Ms}{M_{\ast}}
\newcommand{\Mha}{M_{\rm H{\sc I}}}
\newcommand{\Mhm}{M_{\rm H_2}}
\newcommand{\Mdisk}{M}
\newcommand{\vmax}{v_{\rm max}}
\newcommand{\rflat}{r_{\rm flat}}
\newcommand{\rd}{r_{\rm disk}}
\newcommand{\fg}{f_{\rm atm}}
\newcommand{\dx}{{\rm d}x}
\newcommand{\dr}{{\rm d}r}
\renewcommand{\fd}{f_{\rm M}}
\newcommand{\fj}{f_{\rm j}}
\renewcommand{\d}{{\rm d}}
\newcommand{\Q}{Q_{\rm atm}}
\newcommand{\xc}{x_{\text{core}}}
\newcommand{\xh}{x_{\text{hole}}}
\newcommand{\rh}{r_{\text{hole}}}

\usepackage{color}

\begin{document}

\title{Angular Momentum Regulates Atomic Gas Fractions of Galactic Disks}

\author{D. Obreschkow$^1$}
\author{K. Glazebrook$^2$}
\author{V. Kilborn$^2$}
\author{K. Lutz$^2$}

\affiliation{$^1$International Centre for Radio Astronomy Research (ICRAR), M468, University of Western Australia, WA 6009, Australia}\affiliation{$^2$Centre for Astrophysics and Supercomputing, Swinburne University of Technology, P.O. Box 218, Hawthorn, VIC 3122, Australia}

\begin{abstract}
We show that the mass fraction $\fg=1.35\Mha/\Mdisk$ of neutral atomic gas (\ha and He) in isolated local disk galaxies of baryonic mass $\Mdisk$ is well described by a straightforward stability model for flat exponential disks. In the outer disk parts, where gas at the characteristic dispersion of the warm neutral medium is stable in the sense of Toomre (1964), the disk consists of neutral atomic gas; conversely the inner part where this medium would be Toomre-unstable, is dominated by stars and molecules. Within this model, $\fg$ only depends on a global stability parameter $q\equiv j\sigma/(G\Mdisk)$, where $j$ is the baryonic specific angular momentum of the disk and $\sigma$ the velocity dispersion of the atomic gas. The analytically derived first-order solution $\fg=\min\{1,2.5q^{1.12}\}$ provides a good fit to all plausible rotation curves. This model, with no free parameters, agrees remarkably well ($\pm 0.2$ dex) with measurements of $\fg$ in isolated local disk galaxies, even with galaxies that are extremely \ha-rich or \ha-poor for their mass. The finding that $\fg$ increases monotonically with $q$ for pure stability reasons offers a powerful intuitive explanation for the mean variation of $\fg$ with $\Mdisk$: in a cold dark matter universe galaxies are expected to follow $j\propto\Mdisk^{2/3}$, which implies the average scaling $q\propto\Mdisk^{-1/3}$ and hence $\fg\propto\Mdisk^{-0.37}$, in agreement with observations.
\end{abstract}

\maketitle


\section{Introduction}\label{section_introduction}

Galaxy evolution studies have come a long way since the first detection of atomic hydrogen (\ha) in 21cm-emission in the Milky Way 75 years ago \citep{Ewen1951}. It is now clear that the relative amount of this neutral atomic gas is an elementary property of all galactic disks. Cosmological simulations \citep{Welker2016,Lagos2016,Nelson2015} assert that this gas tells us about a galaxy's past and determines much of its future growth and morphology, even if set for a major galaxy--galaxy collision. The amount of neutral atomic gas is suitably quantified by
\be
	\fg \equiv \frac{1.35\Mha}{\Mdisk},
\ee
where $\Mdisk=\Ms+1.35(\Mha+\Mhm)$ is the disk baryonic mass, $\Ms$ is the stellar mass in the disk, $\Mha$ is the neutral atomic hydrogen (\ha) mass, $\Mhm$ is the molecular hydrogen (\hm) mass and the factor 1.35 accounts for the universal $\sim$26\% helium fraction at redshift $z=0$. 

Observationally, low-mass disks tend to have higher values of $\fg$ than more massive ones \citep{Catinella2013}. In fact, most galaxies with $\Ms<10^{10}\msun$ are \ha dominated \citep{Maddox2014}. 
The reasons for this trend are often deferred to other empirical correlations, such as the observed dependence of the local \hm/\ha ratio on stellar and gas densities and on metallicity \citep[e.g.][]{Blitz2006,Leroy2008,Bolatto2011,Wong2013}. In semi-analytic models \citep{Obreschkow2009b,Lagos2011,Popping2014b} and hydrodynamic simulations (refs.\ above), which aim to reproduce $\fg$ across a representative range of local galaxies, the dominant driver behind $\fg$ is typically hidden in the prescriptions for disk growth and star formation. The intriguing possibility for a more direct explanation is fuelled by the interesting, but unexplained discovery that $\fg$ scales about linearly with $j/M$ \citep{Obreschkow2015a} in local disk galaxies, where $j$ is the baryonic specific angular momentum. Similarly, \cite{Zasov1989} and \cite{Huang2012} found that the \ha fractions in late-type galaxies of similar mass are strongly correlated with $j$ and the spin parameter (a proxy of $j/M^{2/3}$), respectively.

In this Letter, we explain the variation of $\fg$ as a function of $j/M$ using a straightforward model of galactic disks in equilibrium, accounting for all the other galaxy evolution physics only through its overall effect on $j/M$. In \S\ref{section_model} we derive the mathematical model for $\fg$. \S\ref{section_data} then critically compares the model predictions against observations and explains the mean mass dependence of $\fg$.

\section{Analytical model of $\fg$}\label{section_model}

The leading idea of our model is that galactic disks in gravitational equilibrium are made exclusively of neutral atomic gas (\ha and He), if the differential rotation makes this gas stable against gravitational (Jeans) instabilities. In turn, all regions where the atomic gas would be unstable are composed exclusively of molecular and stellar material. This assumption does not imply that the stellar regions remain unstable, since the star formation process can locally re-stabilize the disk by increasing the dispersion and removing gas \citep{Krumholz2012,Governato2010}. The assumption that, in any region, the material is either 0\% or 100\% atomic, is a vast simplification, as all galaxies with resolved maps of stars and \ha show that these components can co-exist in close proximity. However, we will see that this simplification suffices to obtain a good approximation of the \emph{total} atomic gas fraction in a disk galaxy.

An axially symmetric thin disk in gravitational equilibrium is stable if and only if the so-called Toomre parameter $Q$ \citep{Toomre1964} is larger than one. For a disk of neutral atomic gas this parameter reads
\be\label{eq_Qoriginal}
	\Q = \frac{\sigma\kappa}{\pi G\Sigma},
\ee
where $\sigma$ is the local radial velocity dispersion of the atomic gas, $\kappa$ is the local epicyclic frequency, and $G$ is the gravitational constant. A priori, $\Sigma$ is the surface density of the atomic gas. However, since we are only interested in the regions with $\Q>1$, where our model disk is entirely atomic, we can identify $\Sigma$ with the surface density of all disk material. This justifies the assumption that $\Sigma$ follows an exponential profile of scale radius $\rd$,
\be\label{eq_Sigma}
	\Sigma(r) = \frac{M}{2\pi\rd^2} e^{-r/\rd},
\ee
with disk baryonic mass $M=\int_0^\infty 2\pi r~\Sigma(r)\,\dr$. The idea to identify the neutral atomic gas with stable regions of this gas ($\Q>1$) then implies
\be\label{eq_fg_original}
	\fg = \int_{\Q(x)>1} x\,e^{-x}\dx,
\ee
where $x\equiv r/\rd$ is the normalized radius. It is worth noting that the application of Toomre stability to star formation has limitations and caveats \citep[e.g.][]{Elmegreen2011}. Our model is only weakly concerned by these limitations, since it does not depend on the way instabilities develop in a thick, multi-component disk with stellar feedback, but only on the question where the \ha gas, in a thin, single-component disk, is stable.

In what follows, we assume circular orbits, in which case the epicyclic frequency $\kappa$ is related to the radius $r$, the angular velocity $\Omega(r)$ and the velocity $v(r)=r\Omega(r)$ via
\be\label{eq_kappa}
	\kappa = \sqrt{\frac{2\Omega}{r}\,\frac{\d(r^2\Omega)}{\dr}} = \sqrt{\frac{2v}{r^2}\,\frac{\d(rv)}{\d r}}.
\ee

Solving \eq{fg_original} requires an explicit expression for $\Q(x)$, that is expressions for $\sigma(r)$ and $v(r)$. The one-dimensional dispersion $\sigma$ of \ha is set by the stable temperature $T\approx8,000\rm\,K$ of the warm neutral medium (WNM, \citealp{Cox2005}) and shows little variation with galaxy radius. In nearby disks estimates vary from about $8~\kms$ \citep{Shostak1984,Dickey1990} to slightly higher values of $11\pm3~\kms$ \citep{Leroy2008}. We therefore adopt the constant $\sigma=10~\kms$ for this letter, but other choices near this value make no significant changes in the results. 

\subsection{Constant rotation velocity}

\begin{figure}[t]
\begin{center}
\includegraphics[width=\columnwidth]{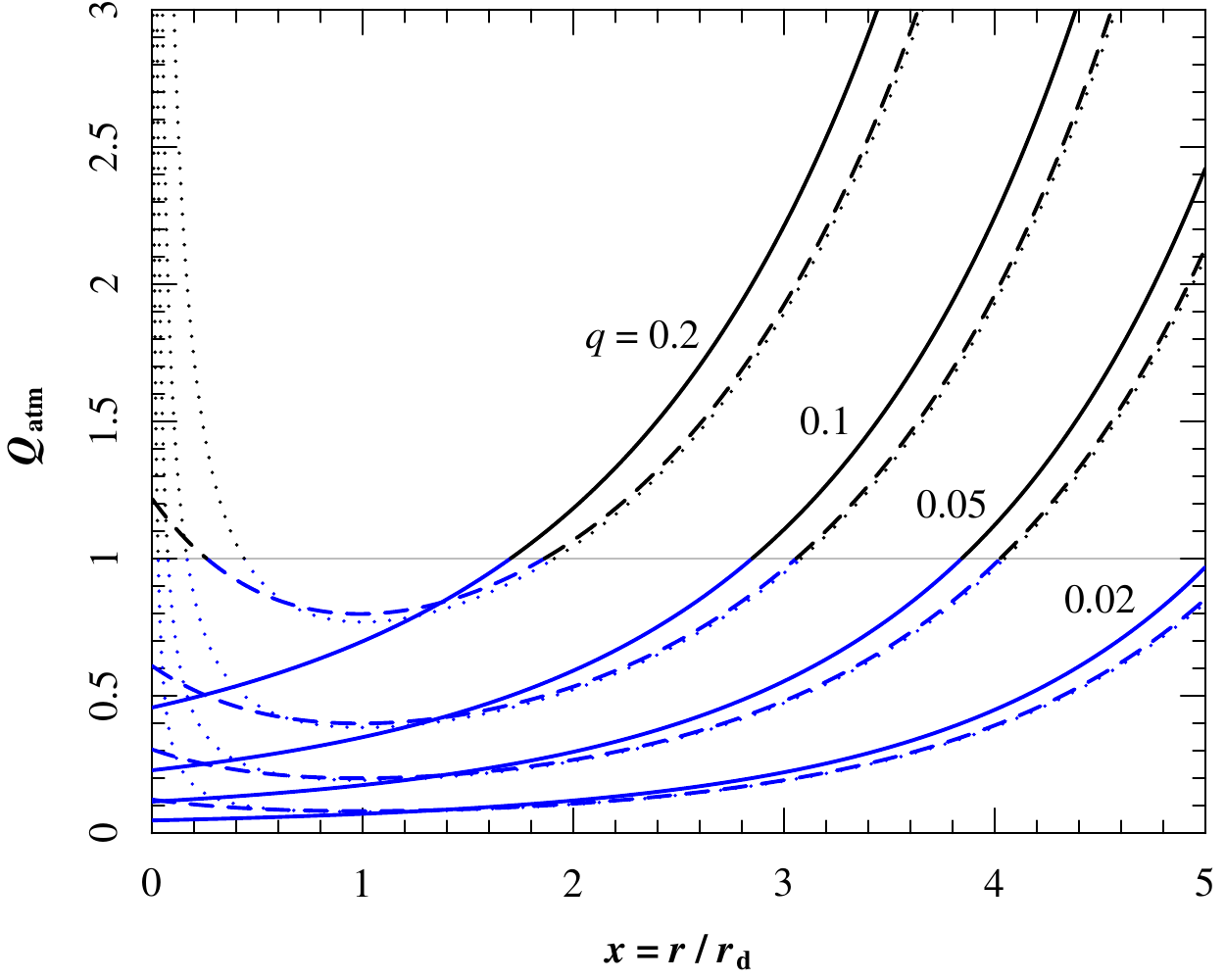}
\caption{Radial variation of the Toomre parameter $\Q$ of neutral atomic gas in the cases of a constant circular velocity (\eq{Q1}, dotted lines) and differential rotation (\eq{Q2}) with $a=1$ (solid lines) and $a=3$ (dashed lines). Only material characterized by $\Q>1$ (black) is assumed to remain atomic.}
\label{fig_q}
\end{center}
\end{figure}

In the case of $v(r)=\vmax~\forall r\geq0$, \eq{kappa} reduces to $\kappa=\sqrt{2}\,\vmax/r$ and hence \eq{Qoriginal} becomes
\be\label{eq_Q1}
	\Q = \sqrt{2}\,q\,x^{-1}e^x.
\ee
Here, $q$ is a dimensionless ``global disk stability parameter,'' originally introduced by \cite{Obreschkow2014} (hereafter OG14), defined as
\be\label{eq_defq}
	q \equiv \frac{j\sigma}{GM},
\ee
where $j\equiv|\mathbf{J}|/M$ is the baryonic specific angular momentum of the disk. In a constantly rotating exponential disk, $j=2\rd\vmax$.

\eq{Q1} is plotted in \fig{q} (dotted lines) for four typical values of $q$. \eq{fg_original} can be solved analytically: if $q\geq 1/(\sqrt{2}e)$, the disk is entirely stable ($\Q\geq1~\forall x\geq0$), hence $\fg(q)=1$; this is a model limitation discussed in \S\ref{subsection_fg_q}. If $q<1/(\sqrt{2}e)$, then $\Q(x)$ intersects the $\Q=1$ line twice (see \fig{q}). The inner and outer intersection points are $\xc=-W_0(-\sqrt{2}q)$ and $\xh=-W_{-1}(-\sqrt{2}q)$, where $W_0$ and $W_{-1}$ are branches of the Lambert $W$ function. As we shall see (\S\ref{subsection_diff_rot}), the core stability zone at $x<\xc$, which only makes a minor mass contribution (up to $26\%$, but often much smaller), is an artifact of assuming a constant rotation velocity. It disappears for most realistically rising rotation curves. We therefore chose to remove this zone from $\fg$, simplifying \eq{fg_original} to $\fg(q) = \int_{\xh}^\infty\,x\,e^{-x}\dx$. Hence,
\be\label{eq_fgq}
	\fg = (1+\xh) e^{-\xh}.
\ee

In the limit $q\rightarrow0^+$, \eq{fgq} yields the Taylor approximation $\fg=\sqrt{2}q+\mathcal{O}(q^2)$, explaining the approximate proportionality between $j/M$ and $\fg$ found in \cite{Obreschkow2015a}. Better first-order approximations are found in log-space. In the vicinity of $q=0.1$, this approximation truncated to $\fg\leq1$ reads
\be\label{eq_fg_approx}
	\fg = \min\{1,2.5q^{1.12}\}.
\ee
This approximation (thick gray line in \fig{fatm_q}) yields errors below 15\% in $\fg$ for $\fg\in(0.01,0.6)$.

An immediate aside from this derivation is the testable prediction for the inner \ha radius. We expect that disks transition from the star+\hm-dominated part to the \ha-dominated part at the radius $\rh=\rd\xh$, where
\be
	\xh = -W_{-1}(-\sqrt{2}q) = -1-W_{-1}(-\fg/e).
\ee
This radius only exists if $q<1/(\sqrt{2}e)$. Otherwise, disks are expected to be \ha-dominated at all radii. We leave the test of this side result to future work.

\subsection{Differential rotation}\label{subsection_diff_rot}

The assumption of a constant rotation velocity $v$ brings the caveat of a diverging epicyclic frequency $\kappa$ as $r\rightarrow0$, making the galaxy unrealistically stable at its center ($x<\xc$). This divergence disappears if we account for the linear increase of the velocity, $v\propto r$, in the center. Such linear behavior is naturally provided by the rotation curve model \citep{Boissier2003},
\be\label{eq_vdiff}
	v(r) = \vmax\Big(1-e^{-r/\rflat}\Big),
\ee
where $\rflat$ is the kinematic radius and $\vmax$ the maximal rotation velocity reached asymptotically as $r\rightarrow\infty$. Note that adopting similar rotation curve models \citep[e.g.][]{Courteau1997} with the same asymptotic behavior leads to very similar results. Our reason for choosing \eq{vdiff} is that, following Equation~(8) in OG14, the specific angular momentum then takes the simple form
\be\label{eq_ja}
	j = 2\rd\vmax\left[1-(1+a)^{-3}\right],
\ee
where $a\equiv\rd/\rflat$. Using this expression for $j$, and substituting \eq{vdiff} into \eq{kappa}, \eq{Qoriginal} becomes
\be\label{eq_Q2}
	\Q = \sqrt{2}qx^{-1}e^x\frac{\sqrt{1-e^{-ax}}\sqrt{1+(ax-1)e^{-ax}}}{1-(1+a)^{-3}}.
\ee
Here, $q$ remains as defined in \eq{defq}, but $j$ is now calculated via \eq{ja}. As expected, \eq{Q2} reduces to \eq{Q1} in the limit of $\rflat\rightarrow0$, i.e.\ $a\rightarrow\infty$, corresponding to a constant rotation. The functions $\Q(x)$ of \eq{Q2} are shown in \fig{q} for $a=1$ and $a=3$. These two values approximate dwarf galaxies and massive spirals, respectively (e.g.~Figure~5 in OG14).

We cannot find explicit closed-form solutions for \eq{fg_original} when $\Q$ is taken from \eq{Q2}, but precise numerical solutions can easily be obtained. The solutions for $a=1$ and $a=3$ are shown in \fig{fatm_q} (solid and dashed black lines). Interestingly, these solutions are very similar to that of a constant rotation curve. The approximate solution of \eq{fg_approx} (gray line in \fig{fatm_q}) remains valid for all rotation curves, making it a universal approximation.

\subsection{Physical interpretation of $q$}

We have seen that  $\fg$ only depends on $q$ but what is its physical interpretation? It turns out that $q$ is a {\it global
analog} to the local Toomre stability parameter, it depends only on global quantities and is a mass-weighted average
of $\Q$ (not the total $Q$). For  $\langle\Q\rangle=\int_0^y\Q\,x\,e^{-x}\dx$ with $y\equiv r_{\rm max}/\rd$, we obtain
\begin{equation*}
	\langle\Q\rangle = \sqrt{2}q\cdot\left\{
	\!\begin{array}{ll}
		y & \text{if $v$ const.}, \\
		\!\frac{\int_0^y\dx\sqrt{1-e^{-ax}}\sqrt{1+(ax-1)e^{-ax}}}{1-(1+a)^{-3}} & \text{otherwise},
	\end{array}
	\right.
\end{equation*}
meaning that $\langle\Q\rangle \propto q$ within any radius $r_{\rm max}$. A variation of this proportionality relation was used as an Ansatz to explain the $j/M$--morphology relation in large disk galaxies (OG14) and to argue that massively star-forming disks owe their clumpy structure to low $j$ more than to high gas fractions \citep{Obreschkow2015b}.


\section{Results and Discussion}\label{section_data}

Let us now compare the model for $\fg$ against empirical data. \S\ref{subsection_fg_q} performs a direct comparison between measurements of $\fg$ and predictions based on observed values of $q$. In \S\ref{subsection_fg_m}, we derive an equation for the mean dependence of $\fg$ on the disk mass $M$, assuming a realistic distribution of angular momenta, and compare this model against data. All stellar masses have been re-derived from $K$-band magnitudes with a uniform mass-to-light ratio of unity, mostly neglecting the contribution (0--$20\%$) from stellar bulges.

\subsection{Variation of $\fg$ with $q$}\label{subsection_fg_q}

\begin{figure}[t]
\begin{center}
\includegraphics[width=\columnwidth]{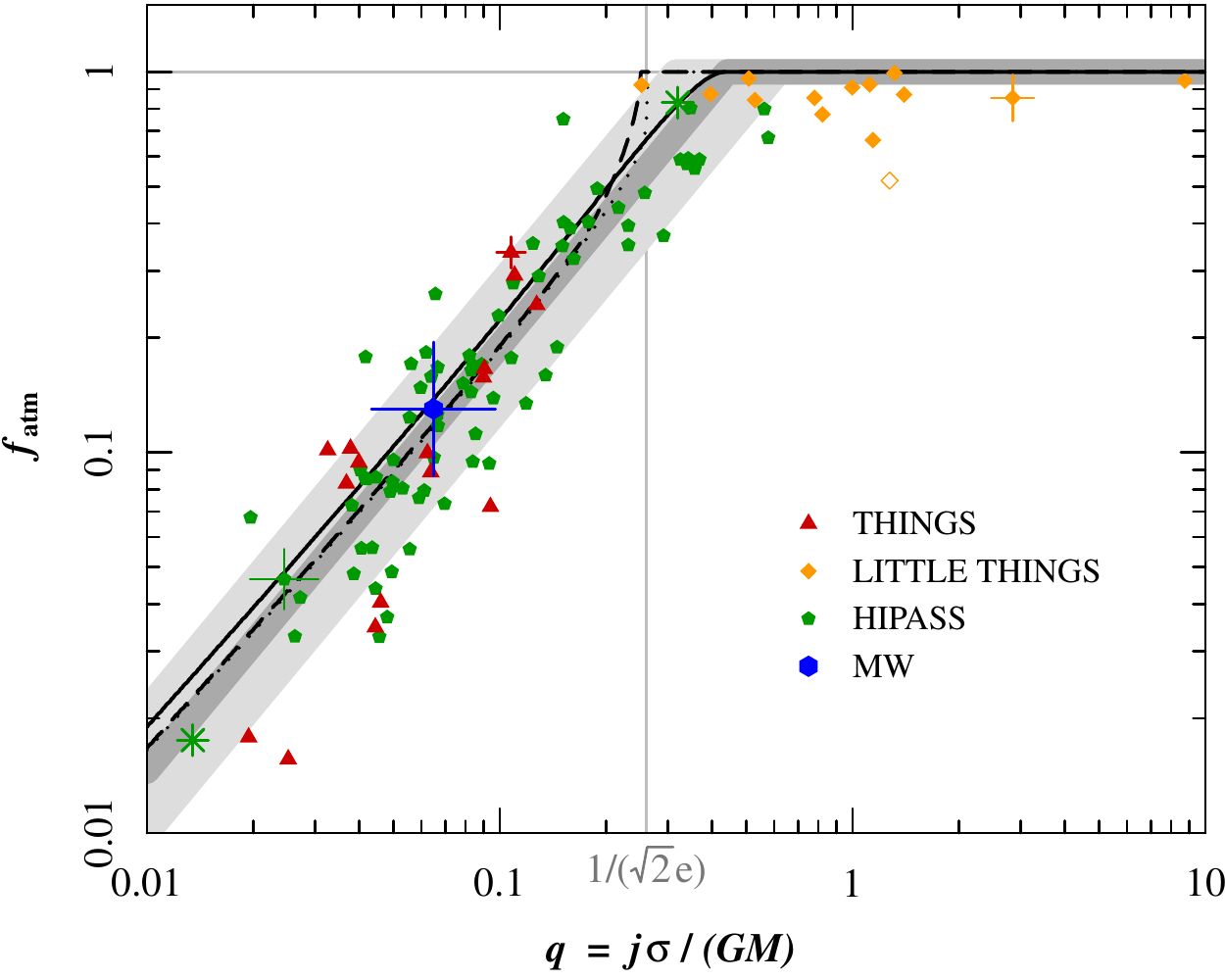}
\caption{Comparison of our model for $\fg$ against data in the local universe. Black lines correspond to different rotation curves from a constant rotation velocity (dotted) to differential rotations with $a=3$ (dashed) and $a=1$ (solid). The dark gray line is the first-order approximation of \eq{fg_approx} with a 40\% scatter shown as light-gray shading. Observational data are shown as colored symbols. For clarity, only some typical error bars are displayed. The empty yellow diamond is the LITTLE THINGS disk with the highest morphological asymmetry. Green stars are HIPASS outliers on the $\fg$-mass relation (\fig{fatm_m}).}
\label{fig_fatm_q}
\end{center}
\end{figure}

\fig{fatm_q} compares the model of $\fg(q)$ to nearby disk galaxies from different surveys. We stress that the model (gray line) is not a fit and has zero free parameters.

Red and yellow points represent the galaxies with the most accurate ($\sim5\%$) measurements of $j$ and $M$, hence the most precise $q$-values. The $j$-measurements rely on a pixel-by-pixel integration of the angular momentum using high-resolution \ha maps as velocity tracers. Red points show the 16 regular spiral galaxies from THINGS \citep{Walter2008}, for which such precise $j$-measurements were obtained by OG14. Yellow points represent dwarf galaxies from LITTLE THINGS \citep{Hunter2012} with $j$-values measured by \cite{Butler2016}. Disk masses $M$ include stellar masses (based on {\it Spitzer} data) and \ha masses (from VLA). The THINGS disk masses also include \hm masses, inferred from CO data. No \hm is included in LITTLE THINGS, but the \hm/\ha ratio is very small ($\ll0.1$) in such low-mass systems \citep{Leroy2005,Schruba2012}.

Green points in \fig{fatm_q} represent galaxies from the HIPASS blind \ha survey \citep{Meyer2004}. Here, we use the subsample (860 galaxies), for which high-fidelity $K$-band magnitudes where compiled by \cite{Meyer2008}. The baryonic disk masses are determined from $K$-band stellar masses and \ha masses. Since \hm masses are not readily available for this sample, we include a uniform \hm fraction ($\Mhm/\Mdisk$) of 4\%, the mean measured in our 16 THINGS galaxies. The $j$-values are approximated as $2\rd\vmax$ (assuming twice the $j$ for \ha and correcting via Eq.~(6) in OG14), where the exponential radius $\rd$ is computed from $r$-band radii in the optical cross-match catalog \citep{Doyle2005} and $\vmax$ is taken to be equal to the \ha half-linewidth, corrected for the optical inclination. The galaxies processed in this way exhibit a relatively large ($\sim0.35~\rm dex$) scatter around our model of $\fg(q)$, consistent with the expected large empirical uncertainties. In \fig{fatm_q} we therefore limited this sample to a more accurate subsample of 75 objects with certain optical-\ha crossmatching (no source confusion), reliable inclination corrections (edge-on galaxies with inclinations larger than 60 degrees), reliable Hubble-flow distances ($>10~\rm Mpc$) and a large number of optical resolution elements (effective radius $>20\arcsec$).

Finally, the blue diamond shows the Milky Way. Here, $\fg$ was taken from \cite{Obreschkow2011c} and $j=2\rd\vmax$ with $\vmax=220\,\kms$ and $\rd=3.9\rm\,kpc$ for \emph{all} disk baryons (\citealp{Kafle2014} find 4.92\,kpc for a Miyamoto-Nagai disk, which must be multiplied by 0.82 for an exponential disk).

Overall, \fig{fatm_q} demonstrates a remarkable agreement between model and observations. The Spearman correlation between $q$ and $\fg$ is 0.91, and the rms offset between data and model is about $0.2~\rm dex$ in $\fg$. This scatter is better than the scatter ($\sim0.3\rm~dex$) of current empirical predictors of $\fg$ based on optical data \citep{Catinella2012}. If we account for the fact that the \ha dispersion $\sigma$ exhibits some intrinsic empirical scatter of about 40\% (light gray zone in \fig{fatm_q}), data and model become statistically consistent. Moreover, the offset of the data is virtually uncorrelated with mass $M$ (Pearson coefficient below 0.1), making it clear that the entire mass dependence of $\fg$ must be contained in the mass dependence of $q$ -- an idea developed in \S\ref{subsection_fg_m}.

We performed a critical analysis of the parameter $q$ as a predictor of $\fg$, relative to other candidates. Explicitly, we checked if the overall rotational support, quantified by $\sigma/\vmax$ is a suitable predictor of $\fg$. We also investigated a model where the disk baryons are atomic if and only if the surface density lies below a threshold $\Sigma_{\rm crit}$. Both models produce more scatter ($>0.3\rm~dex$) and lower Spearman ($\sim0.7$ with clear outliers) than $\fg(q)$ in \fig{fatm_q}.

The clearest limitation of our model is its prediction that galaxies with $q>1/(\sqrt{2}e)$, typically small dwarf galaxies, are fully atomic ($\fg=1$). In reality, such galaxies are never strictly dark, although their baryon mass is dominated by \ha \citep{Doyle2005}. This limitation originates directly from the assumption that there are no partially atomic regions in the disk.

\begin{figure}[t]
\begin{center}
\includegraphics[width=\columnwidth]{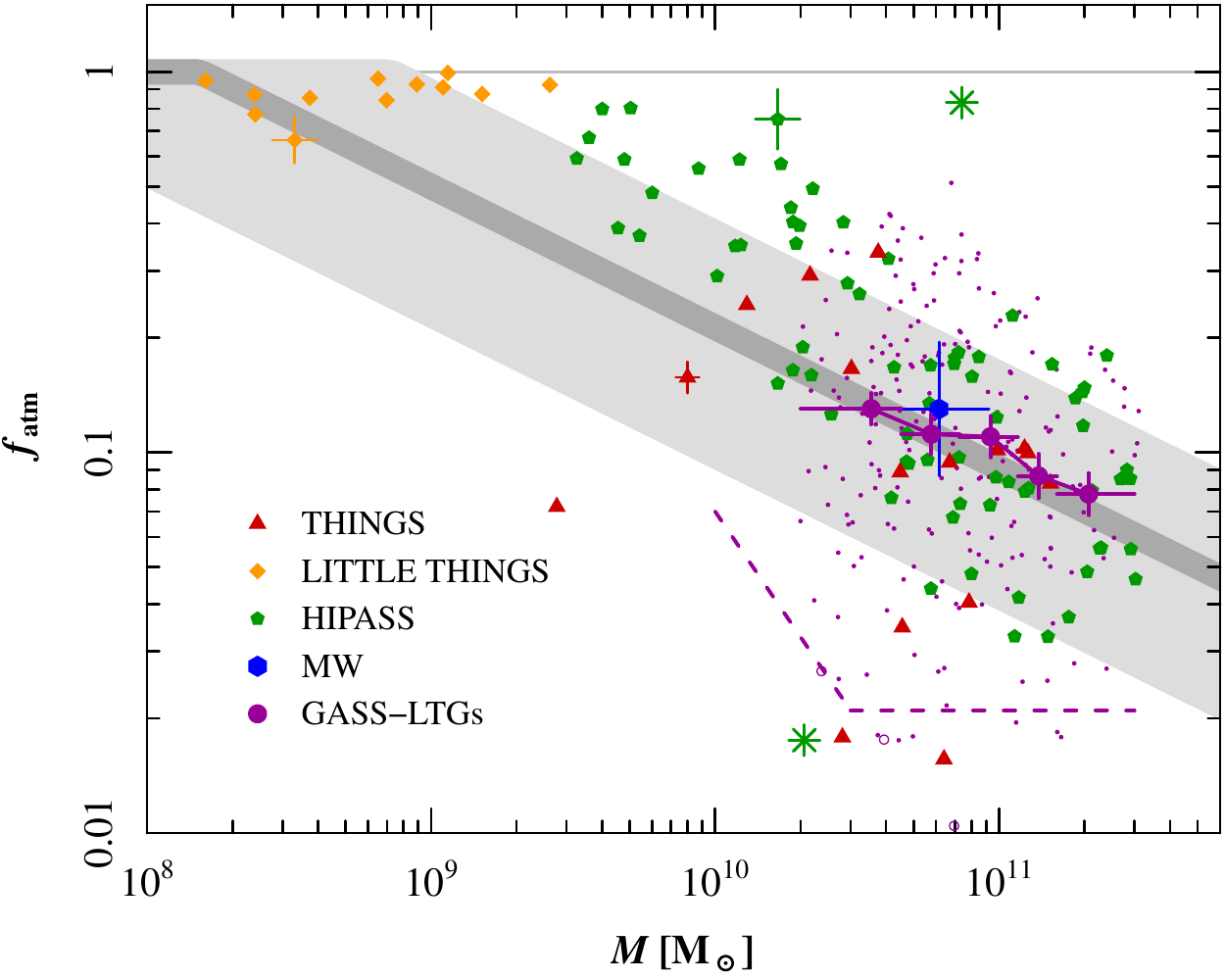}
\caption{Atomic gas fraction versus baryonic disk mass. Different symbols represent the same data sets as in \fig{fatm_q}. Small purple points are the disk galaxies of the representative sample of the GASS survey with upper limits for non-detections shown as empty circles. The dashed line delimits the approximate detection limit of GASS. The large points are geometric averages in mass bins (including the upper limits). The thick gray line represents our linear model of \eq{fg_m_mean}. Shading indicates the 68\%-scatter around this relation, expected from the log-normal spin parameter distribution.}
\label{fig_fatm_m}
\end{center}
\end{figure}

\subsection{Variation of $\fg$ with $M$}\label{subsection_fg_m}

\fig{fatm_m} shows $\fg$ as a function of the disk mass $M$ for the same galaxies as in \fig{fatm_q}. In addition it includes data from the `representative sample' of 760 galaxies at $z\approx0$, published with the final data release (DR3) of the GASS survey \citep{Catinella2010,Catinella2013}. At stellar masses $\Ms>10^{10}\msun$ in this sample, the true distribution of \ha fractions ($\gtrsim2\%$) is sampled uniformly. We obtained $K$-band magnitudes from 2MASS \citep{Jarrett2000} and selected the disk-dominated galaxies roughly by retaining only concentration parameters $r_{90}/r_{50}$ below 2.8 \citep{Park2005} in $r$-band and excluding galaxies in the red sequence (NUV--$r>4.5$) of the color-magnitude diagram (Fig.~4 in \citealp{Catinella2013}). The resulting sample counts 213 galaxies (including 3 \ha-non-detections). It is the only sample in \fig{fatm_m} revealing an unbiased mean relation between $\fg$ and $M$. As in our HIPASS data, we included a constant 4\% \hm fraction in $M$ with negligible implications.

The scatter in \fig{fatm_m} at fixed $M$ is about twice as large as in \fig{fatm_q} at fixed $q$. Moreover, in \fig{fatm_m}, different datasets are offset to each other in $\fg$, whereas they occupy the same locus in \fig{fatm_q}, implying that $\fg(q)$ is insensitive to \ha selection bias. Particularly interesting are the two galaxies from HIPASS marked as green stars in Figures~\ref{fig_fatm_q} and \ref{fig_fatm_m}. These objects are field galaxies that are extremely \ha-poor and \ha-rich for their mass; however their \ha fraction seems quite normal for their $q$-values. In other words, these galaxies are simply outliers in their relative amount of angular momentum, probably explainable by their (past) cosmic environment. These results confirm that $q$ is fundamental for setting $\fg$ in disk galaxies.

To explain the mean trend of $\fg$ with $M$ in our framework, we need to express $q$ in $\fg(q)$ as a function of $M$. Assuming that disk galaxies condense out of scale-free cold dark matter halos we expect galactic disks to fall on a relation $j=k M^{2/3}$ (OG14; \citealp{Romanowsky2012}), where $k$ depends on the halo spin parameter $\lambda$ \citep{Bullock2001b} and the fractions of retained mass $\fd=M/M_{\rm halo}$ and specific angular momentum $\fj=j/j_{\rm halo}$. For simplicity we will assume $\fd=0.05$ and $\fj=1$, typical for Milky Way--like disks, although in general one should vary both to account for the physics of galaxy formation. Nevertheless, we get a remarkably good agreement for the general trend. The halo spin parameters exhibit a skewed distribution of mode $\lambda\approx0.03$ \citep{Bullock2001b}, leading to $k\approx90$ for $j$ in units of $\rm kpc~\kms$ and $M$ in units of $10^9\msun$. Substituted into \eq{defq}, we obtain
%
\be\label{eq_trend_q}
	q = \frac{k\Mdisk^{2/3}\sigma}{G\Mdisk} \approx 0.22\left(\frac{M}{10^9\msun}\right)^{-1/3}.
\ee
%
for $\sigma=10~\kms$ (\S\ref{section_model}). With \eq{fg_approx}, we then find the approximate mean trend
\be\label{eq_fg_m_mean}
	\fg \approx 0.5\left(\frac{M}{10^9\msun}\right)^{-0.37},
\ee
as long as $\fg<1$. This function is shown as thick line in \fig{fatm_m} with the gray shading representing the large 68\% quantile scatter expected from the log-normal $\lambda$ distribution \citep{Bullock2001b}. The agreement of this model with the empirical data is good. All empirical data points, except for those from the GASS survey (purple), come from \ha-selected surveys and therefore naturally lie above the mean expectation of $\fg(M)$.


\section{Conclusions}\label{section_conclusion}

A straightforward model, assuming that neutral atomic gas in disk galaxies remains atomic if and only if it is Toomre-stable at the typical temperature of the WNM, can explain the observed gas fractions in nearby star-forming galaxies. This model only uses the new global stability parameter $q=j\sigma/(G\Mdisk)$, which depends on the baryonic disk mass $M$, the specific angular momentum of the disk $j$ and a universal dispersion $\sigma=10\,\kms$ (characteristic of the WNM) to predict the observed values of $\fg$, without any free parameters. The predictions are accurate within about $50\%$, over at least four orders of magnitude in $M$, three orders of magnitude in $j$ and two orders of magnitude in $\fg$. A comparison to the less fundamental but more common representation of $\fg$ as a function of $M$ demonstrates that the specific angular momentum $j$ is an indispensable quantity in understanding $\fg$. In turn, if we model $\fg$ as a function of $q$, there is virtually no residual dependence on $M$. The mean trend of $\fg$ as a function of $M$ is then explainable from the distribution of disk galaxies in the ($M$,$j$)-plane.

These findings suggest that $\fg$ depends on a galaxy's assembly history and past interactions only (or at least mainly) via the history dependence of $q\propto j/M$, an interesting test case for hydrodynamic simulations of galaxy formation. The presented results focus on isolated disk galaxies, excluding systems in cluster environments and early-type galaxies. Studying the deviations from this model as a function of morphology and environment are interesting avenues for future research.


\section*{Acknowledgements}

This Letter was supported by the Australian Research Council (Discovery Project 160102235). We thank Barbara Catinella, Prajwal Kafle, Adam Stevens, and Bruce Elmegreen for helpful input and the THINGS, LITTLE THINGS, HIPASS, and GASS teams for making their survey data publicly available.



\end{document}